\documentclass[conference]{IEEEtran}
\usepackage{amsmath,amsthm}
\usepackage{cite}
\usepackage{graphicx}
\usepackage{epstopdf}
\usepackage{amsfonts,amsmath,amssymb}
\usepackage{cite}
\usepackage{graphicx}
\usepackage{url}
\usepackage{bm}
\usepackage{bbm}
\usepackage{amssymb}

\begin{document}

\title{{Quantum Key Distribution over Combined Atmospheric Fading Channels}}
\author{
\IEEEauthorblockN{
Nedasadat Hosseinidehaj and Robert Malaney}
\IEEEauthorblockA{School of Electrical Engineering  \& Telecommunications,\\
The University of New South Wales,
Sydney, NSW 2052, Australia.\\
neda.hosseini@student.unsw.edu.au, r.malaney@unsw.edu.au}
}

\vspace{-5cm}

\maketitle
\begin{abstract}
In this work we analyze a quantum communication scheme for  entanglement-based continuous variable  quantum key distribution  between two ground stations. Communication occurs via a satellite over two independent atmospheric fading channels dominated by turbulence-induced beam wander. In this scheme the engineering complexity remains largely on the ground transceivers, with the satellite acting simply as a reflector. We show how the use of a highly selective post-selection strategy may lead to a useful quantum key generation rate for this system. This work represents the first quantitative assessment of continuous variable quantum key rates in the pragmatic scenario of reflection off low-earth-orbit satellites.

\end{abstract}

\section{Introduction}

Quantum key distribution (QKD) \cite{84} is the most developed and most widely known protocol of quantum communications. A QKD protocol is consists of two steps. Firstly, a quantum communication part where two distant parties, Alice and Bob, generate two sets of correlated data through the exchange of a significant number of quantum states. Secondly, by running a classical post-processing protocol through a public (but authenticated)  classical channel Alice and Bob extract from their correlated data a secret key that is unknown to a potential eavesdropper, Eve. The final key which is unconditionally secure based on the laws of quantum mechanics  can then be used to encode secret messages e.g., \cite{84, G, 3}.

There are two main technologies of QKD, discrete variable (DV) where key information is encoded on the properties of single photons such as the phase or  polarization e.g., \cite{f1, f2}, and continuous variable (CV) where key information is encoded on the quadrature variables of coherent or squeezed states e.g., \cite{1, 3, P, RR, RR2}. In the former technology detection is realized by single photon counting measurements, which are replaced in CV QKD protocols by the homodyne (or heterodyne) detection techniques which are faster and more efficient.

Although QKD has matured to commercial applications  and a number of QKD schemes have been implemented both over optical fibers \cite{3, G} and terrestrial free-space links \cite{f1, f2}, it is still limited to relatively small scales. One way of extending the deployment range of QKD is through the use of
satellites. Indeed, it is now a widely held view in the quantum communications community that the use of satellites is pivotal the  deployment of quantum based communication protocols over global scales \cite{r7,r8,r9,r10,r11,r12,r13,r14,r15,r16,r17}. Such satellite-based quantum communication will be built on the techniques of  free-space optical (FSO) communications (for review see \cite{fso}).
Implementations of QKD over atmospheric channels are  discussed in several recent works \cite{r10, pp1, pp2, pp3, 21}. All of the free-space QKD systems (DVs or CVs) so far implemented are based on direct transmission through a single point-to-point free-space link. In this work we will focus on  CV QKD protocols over the combined atmospheric fading channel traversed by a laser beam reflected off a low earth orbit (LEO) satellite.

The main motivation for our scheme,  referred to as the \emph{direct QKD scheme}, is to minimize the deployment of quantum technology at the satellite.  There are  many practical advantages in deploying quantum aspects of the communication technology at the ground stations, such as lower-cost maintenance and the ability to rapidly upgrade. The deployment likelihood for the type of (relative) low-complexity communication scheme we describe here is enhanced by recent experimental tests of the reflection paradigm for single photons  \cite{r16,r17}). Although satellite reflection towards another station is a sophisticated engineering task in its own right, it does not require on-board generation of quantum communication information and is devoid of any embedded quantum control mechanisms. Our scheme therefore represents one of the simplest ways of creating QKD via satellite. The cost of this simplicity will be a reduction in the  secret key rate, and it is this point that forms the thrust of the work reported here.



As we discuss later, all  QKD schemes can be represented by an equivalent
entanglement-based  QKD protocol. We will focus on CV QKD entanglement-based protocols, where the entangled states shared by the two ground stations are first generated via quantum communication. Specifically we assume a  two-mode squeezed  state  is generated at ground station A, with one component of the beam kept at A, while the other component is transmitted to ground station B via a LEO reflecting relay satellite. The level of entanglement produced by this scheme has recently been analyzed by us in \cite{neda1}. Quantum key generation can then occur via Gaussian measurements e.g., heterodyne or homodyne detection on the components at each ground station \cite{1, P}.
 Note that the transmitted beam from ground station A will encounter atmospheric fading caused by its traversal in the uplink towards the satellite, and then again on its traversal in the downlink towards ground station B.  The fading experienced will be largely  dominated by the transmission fluctuations caused by beam wandering \cite{fso,20,21}.

\section{QKD over fading channels}

\begin{figure}[!t]
    \begin{center}
   {\includegraphics[width=3.2 in, height=2.3 in]{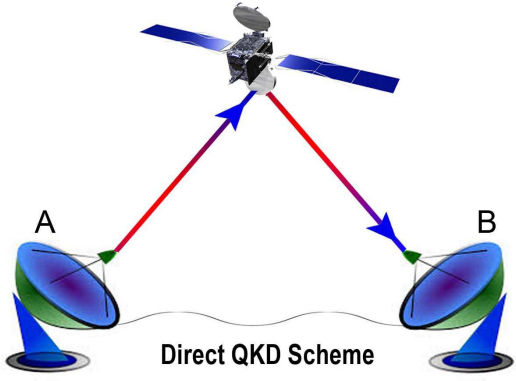}}
    \caption{Direct transmission scheme for implementation of CV QKD. Here one mode of an entangled state remains at A while the other mode is reflected off a LEO satellite and sent to B.  Stations A and B can make a range of quantum measurements. Classical post-selection can occur at B, and classical post-processing between A and B occurs before a quantum key is generated. }\label{fig:scheme}
    \end{center}
\end{figure}

We wish to analyze the direct QKD scheme of Fig.~\ref{fig:scheme}. But let us first introduce some preliminaries regarding CV quantum information. In the following we set $\hbar=2$.

\subsection{Preliminaries}

The quadrature operators $\hat q,\hat p$  for a single bosonic mode are defined by
$\hat q = \hat a + \,{\hat a^\dag }\,,\,\,\,\,\,\hat p = i({\hat a^\dag } - \hat a\,)$
where  $\hat a,\,{\hat a^\dag }$ are the annihilation and creation operators, repectively.
 The quadratures satisfy the commutation relation $\left[ {\hat q,\,\hat p} \right] = 2i$.
 The vector of quadrature operators for a quantum state with $n$ modes can then be defined as
${\hat R_{1, \ldots ,n}} = \left( {{{\hat q}_1},\,{{\hat p}_1}, \ldots ,{{\hat q}_n},{{\hat p}_n}\,} \right)$.

We note that as opposed to some of the non-Gaussian states we discuss later, Gaussian states are characterized solely by the first and second moments of the  quadrature operators. These second moments can be represented by a  \emph{covariance matrix} (CM) $M$, whose elements are given by
\begin{eqnarray}\label{ad1}
{M_{ij}} = \frac{1}{2}\left\langle {{{\hat R}_i}{{\hat R}_j} + {{\hat R}_j}{{\hat R}_i}} \right\rangle  - \left\langle {{{\hat R}_i}} \right\rangle \left\langle {{{\hat R}_j}} \right\rangle .
\end{eqnarray}
The CM of a $n$-mode quantum state is a $2n \times 2n$ real and symmetric matrix which must satisfy the uncertainty principle, \emph{viz.},
$M  + i\,\Omega \, \ge \,0$,
where
\begin{eqnarray}\label{ad2}
\Omega : = \mathop  \oplus \limits_{k = 1}^n \,\,\omega  = \left( {\begin{array}{*{20}{c}}
\omega &{}&{}\\
{}& \ddots &{}\\
{}&{}&\omega
\end{array}} \right)\,,\,\,\omega : = \left( {\begin{array}{*{20}{c}}
0&1\\
{ - 1}&0
\end{array}} \right).
\end{eqnarray}
The first moment of every two-mode Gaussian state can be set to zero (By local unitary operators) and the CM can take the following standard form
\begin{equation}
{M_s} = \left( {\begin{array}{*{20}{c}}
A_m&C_m\\
{{C_m^T}}&B_m
\end{array}} \right),\,
\label{pp3}
\end {equation}
where $A_m = aI\,,\,B_m = bI\,,\, C_m = diag\left( {{c_ + },{c_ - }} \right)$,
 $a,b,{c_ + },{c_ - } \in \mathbb{R}$, and
 $I$ is a $2 \times 2$ identity matrix.

\subsection{Quantum Key Rates}

 CV QKD  protocols can be described as prepare-and-measure schemes (which have an equivalent representation as CV entangled protocols), where Alice prepares quantum states based on encoding (modulation) of classical random variables onto Gaussian states, such as squeezed states or coherent states, and then sends them to Bob. For each incoming state, Bob makes Gaussian measurements, e.g., homodyne or heterodyne detection on the amplitude or phase quadrature. In order to warrant security, Alice and Bob must randomly choose different basis for preparation and measurement. When the quantum communication is finished and all the incoming states are measured by Bob, the second stage, i.e., classical post-processing over a public channel starts where Alice and Bob reveal which quadrature (basis) they used to prepare and measure the information. At the second step, the two parties reveal a randomly chosen subset of their data, allowing them to estimate some parameters of the channel and upper bound the information Eve can have about their values. This step is followed by a reconciliation protocol which encompasses error correction (e.g., via LDPC codes \cite{LDPC} combined with digitization). QKD can be operated in two reconciliation scenarios, direct reconciliation (DR) and reverse reconciliation (RR). 
In the DR protocol Alice's data are the reference and she sends correction information (classical information) to Bob who corrects his key elements to have the same values as Alice. By contrast, in RR protocol Bob's data are the reference and must be estimated by Alice (also by Eve) \cite{RR}. Finally, both parties knowing the upper bound on Eve's information, and then apply a privacy amplification protocol to produce a shared binary secret key.

Considering the type of quantum state (squeezed states or coherent states) which Alice prepares and also the kind of measurement (homodyne or heterodyne detection) which Bob applies on the received states as well as the type of reconciliation, there are eight protocols to represent CV QKD in the prepare-and-measure paradigm. However, all the protocols can be described in an unified way using an entanglement-based scheme \cite{1, P}, where Alice and Bob share a two-mode squeezed state $AB$, and they both make a \emph{generalized} \emph{heterodyne} detection on their own modes using an unbalanced beam splitter of transmittivity ${T_A}$ in Alice's side and of transmittivity ${T_B}$ in Bob's side. If Alice applies a homodyne detection (${T_A}=1$), the (equivalent) prepared state is a squeezed state and if Alice makes a heterodyne detection (${T_A}=1/2$), the (equivalent) prepared state is a coherent state. On the other side, Bob can make homodyne measurement with ${T_B}=1$ and heterodyne measurement with ${T_B}=1/2$.



Let us now recall briefly how security is analyzed in the Gaussian CV QKD protocols we investigate. In this paper, the Gaussian entanglement-based scheme for CV QKD  is considered, in which Alice generates an entangled state (pair $AB$) with quadrature variance $v$ of each of its modes. One mode of an entangled state (mode $A$) is kept and measured by Alice (homodyne or heterodyne) while the other mode (mode $B$) is sent through the lossy channel with transmittivity of $\tau$ and measured by Bob using a homodyne detection. At the output of the channel, the entangled quantum state before Alice and Bob's measurements is a Gaussian two-mode state with a zero mean and  the CM of \eqref{pp3} taking the specific form
\begin{equation}
\begin{array}{l}
{M_{AB}} = \left( {\begin{array}{*{20}{c}}
{aI}&{cZ}\\
{cZ}&{bI}
\end{array}} \right)\,, \ {\rm where}
\\ \\
a = v\,,\,\,b = 1 + \tau \left( {v - 1} \right)\,\,,\,\,c = \sqrt {\tau \left( {{v^2} - 1} \right)}
\end{array}
\label{k1}
\end{equation}
where $Z = diag\left( {1, - 1} \right)$. Considering collective eavesdropping attacks (where Eve interacts individually with each signal pulse sent by Alice and applies a joint measurement at the end of the classical post-processing), the secret key rate  $K$ (bits per pulse) in the case for the RR and DR scenarios can be derived.  We point out that due to the non-Gaussian nature of our final ensembles the key rates provided here can be considered lower bounds\footnote{Due to the relatively long coherence time of the channel, in principal it should be possible to devise a scheme in which key rates for each realization of the fading  (each fading bin realized) are derived and summed.  Within each (small) bin we can assume the fading is constant and therefore the states in that particular bin are Gaussian. We will not pursue this type of scheme in this work.}\cite{Gauss1},
 but only on the usual assumption  that the number of exchanges between Alice and Bob are considered infinite (see later). More details on the derivation of $K$ can be found in \cite{1, P}. Here, we summarize these known results for  three specific protocols that we later simulate.

 \emph{(i) Reverse Reconciliation (homodyne by Alice):} For this type of reconciliation we find
\begin{eqnarray}\label{k2}
K = {I_{AB}} - {\chi _{BE}}
\end{eqnarray}
where ${I_{AB}}$ is the mutual information between Alice and Bob  expressed in terms of the quadrature variance and conditional quadrature variance of modes $A$ and $B$, i.e. ${V_A}$ and ${V_{A\left| B \right.}}$ (variance of A conditioned on measurement of B) as
\begin{eqnarray}\label{k9}
{I_{AB}} = \frac{1}{2}\,{{\log }_2} \left( {\frac{{{V_A}}}{{{V_{A\left| B \right.}}}}} \right)
\end{eqnarray}
where ${V_A} = a\,,\,\,{V_{A\left| B \right.}} = a - \frac{{{c^2}}}{b}$.
Eve's quantum information on Bob's measurement can be calculated as
\begin{eqnarray}\label{k3}
{\chi _{BE}} = S(E) - S(E\left| B \right.)
\end{eqnarray}
where $S(E)$ and $S(E\left| B \right.)$ are the von Neumann entropy of  Eve's state before the measurement on mode $B$ and the von Neumann entropy of Eve's state conditioned on the measurement outcome, respectively. Using the fact that Eve's system is able to purify the state $AB$, we will have $S(E) = S(AB)$, where $S(AB)$ can be calculated through the symplectic eigenvalues ${\nu _{1,2}}$ of ${M_{AB}} $ as:
\begin{eqnarray}\label{k4}
S\left( {AB} \right) = G\left( {\frac{{{\nu _1} - 1}}{2}} \right) + G\left( {\frac{{{\nu _2} - 1}}{2}} \right)
\end{eqnarray}
where $G\left( x \right) = \left( {x + 1} \right)\,{{\log }_2} \left( {x + 1} \right)\, - \,x\,{{\log }_2} x$ is the bosonic entropic function. The symplectic eigenvalues ${\nu _{1,2}}$ of ${M_{AB}} $ are provided by
\begin{eqnarray}\label{k5}
{\nu _{1,2}} = \sqrt {\frac{{\Delta \pm \sqrt {{{\Delta}^2} - 4\det {M_{AB}} } }}{2}}
\end{eqnarray}
with $\Delta  = \det A_m + \det B_m + 2\det C_m$. Next, the entropy $S(E\left| B \right.)$ as a function of the symplectic eigenvalue ${\nu _3}$ of the conditional covariance matrix ${M _{E\left| B \right.}}$ is given by
\begin{eqnarray}\label{k6}
S\left( {E\left| B \right.} \right) = G\left( {\frac{{{\nu _3} - 1}}{2}} \right)
\end{eqnarray}
where
\begin{eqnarray}\label{k7}
\begin{array}{l}
{M_{E\left| B \right.}} = A_m - C_m{\left( {\Pi B_m\Pi } \right)^{ - 1}}{C_m^T}\\
\\
\nu _3^2 = a\left( {a - {{{c^2}} \mathord{\left/
 {\vphantom {{{c^2}} b}} \right.
 \kern-\nulldelimiterspace} b}} \right)
\end{array}
\end{eqnarray}
where $\Pi : = diag\left\{ {1,0} \right\}$ and ${\left( {\Pi B_m\Pi } \right)^{ - 1}}$ is a pseudo-inverse since ${\Pi B_m\Pi }$ is singular.

\emph{(ii) Direct Reconciliation (homodyne by Alice):} Alice and Bob's mutual information is the same for DR and RR. However, in DR, Eve's information on Alice's measurement should be calculated by ${\chi _{AE}} = S(E) - S(E\left| A \right.)$,
where $S(E)$ is exactly the same as RR in \eqref{k4}. The conditional entropy $S(E\left| A \right.)$ is also calculated in a similar way as
\begin{eqnarray}\label{k9a}
S\left( {E\left| A \right.} \right) = G\left( {\frac{{{\nu _3} - 1}}{2}} \right)
\end{eqnarray}
where $\nu _3^2 = b\left( {b - {{{c^2}} \mathord{\left/
 {\vphantom {{{c^2}} a}} \right.
 \kern-\nulldelimiterspace} a}} \right)$.
Thus, the key rate for DR can be estimated by $K = {I_{AB}} - {\chi _{AE}}$.

 \emph{(iii) Reverse Reconciliation (heterodyne by Alice)}: In the above discussion on RR and DR we have assumed homodyne detection at Alice.  It will be useful for us to consider a twist on the RR protocol, where Alice makes a heterodyne detection.
When Alice makes such a heterodyne detection on her own mode, the mutual information between Alice and Bob changes such that
\begin{eqnarray}\label{k12}
{I_{AB}} = \frac{1}{2}\,{{\log }_2} \left( {\frac{{{V_A} + 1}}{{{V_{A\left| B \right.}} + 1}}} \right)
\end{eqnarray}
Note that Eve's information on Bob's measurement in the RR scenario is exactly the same as \eqref{k3}.

 \subsection{Atmospheric Turbulence}
 Beam wander is expected to dominate losses in a wide range of turbulent atmospheric channels and is considered to be the dominant loss mechanism in ground-to-satellite channels  \cite{fso,20,21}.  If we assume the beam spatially fluctuates around the receiver's  center point, such fading can be described by a distribution of transmission coefficients $\eta$ with a probability density distribution $p(\eta)$, where this latter function is given by the log-negative Weibull distribution \cite{20} \cite{21},
\begin{equation}\
p\left( \eta  \right) = \frac{{2{L^2}}}{{\sigma _b^2\lambda \eta }}{\left( {2\ln \frac{{{\eta _0}}}{\eta }} \right)^{\left( {\frac{2}{\lambda }} \right) - 1}}\exp \left( { - \frac{{{L^2}}}{{2\sigma _b^2}}{{\left( {2\ln \frac{{{\eta _0}}}{\eta }} \right)}^{\left( {\frac{2}{\lambda }} \right)}}} \right)
\label{f1}
\end{equation}
for $\eta  \in \left[ {0,\,{\eta _0}} \right]$, with $p\left( \eta  \right) = 0$ otherwise.
Here, ${\sigma _b}^2$ is the beam wander variance,
 $\lambda$ is the shape parameter,  $L$ is the scale parameter, and ${\eta _0}$ is the  maximum transmission value. The latter three parameters are given by
\begin{equation}
\begin{array}{c}
\lambda  = 8h\frac{{\exp \left( { - 4h} \right){I_1}\left[ {4h} \right]}}{{1 - \exp \left( { - 4h} \right){I_0}\left[ {4h} \right]}}{\left[ {\ln \left( {\frac{{2\eta _0^2}}{{1 - \exp \left( { - 4h} \right){I_0}\left[ {4h} \right]}}} \right)} \right]^{ - 1}}\\
\\
L = \beta {\left[ {\ln \left( {\frac{{2\eta _0^2}}{{1 - \exp \left( { - 4h} \right){I_0}\left[ {4h} \right]}}} \right)} \right]^{ - \left( {{1 \mathord{\left/
 {\vphantom {1 \lambda }} \right.
 \kern-\nulldelimiterspace} \lambda }} \right)}}\,,\,\,\,\eta _0^2 = 1 - \exp \left( { - 2h} \right)
\end{array}
\label{f2}
 \end{equation}
where ${I_0}\left[ . \right]$ and ${I_1}\left[ . \right]$ are the modified Bessel functions, and where $h = {\left( {{\beta \mathord{\left/
 {\vphantom {a W}} \right.
 \kern-\nulldelimiterspace} W}} \right)^2}$, with $\beta$ being the aperture radius and  $W$  the beam-spot radius.


Note that  the beam wander variance $\sigma _b^2$ for the uplink is normally significantly larger than the downlink due to the fact that turbulence is larger near the ground\cite{fso}. Also note the rate of the fluctuations caused by turbulence is normally much slower than than transmission rates of the light pulses (kHz  compared to Mhz). This allows for measurements of the channel transmission coefficient (using intertwined coherent pulses) to be made dynamically by a ground receiver with the measured classical information being fed back to the sending station, all well within the coherence time of the channel.

\subsection{Direct QKD Scheme}
In the direct transmission scheme, we assume Alice is located at the ground station A and Bob is placed at the station B. Since security analysis, and the subsequent key rate, of the Gaussian CV QKD protocols is based on the CM description of the quantum states, we are required to calculate the CM of the output state of our scheme between the terrestrial stations. Let us consider the ground station A initially possessing a two-mode squeezed vacuum state with squeezing $r$, then the initial CM can be written
\begin{eqnarray}\label{D1}
{{M}_i} = \left( {\begin{array}{*{20}{c}}
{v\,I}&{\sqrt {{v^2} - 1} \,Z}\\
{\sqrt {{v^2} - 1} \,Z}&{v\,I}
\end{array}} \right) ,
\end{eqnarray}
where $v = \cosh \left( {2r} \right)$,  $r \in \left[ {0,\,\infty } \right)$. We assume one mode remains at the ground station while the other mode is transmitted over the fading uplink to the satellite, then perfectly reflected in the satellite and sent through the fading downlink toward the ground station B. As a result, depending on the initial level of squeezing, there would exist an entangled state between the two ground stations. The separate uplink and downlink channels are assumed to be independent and non-identical.

After transmission of the optical mode through the uplink and then reflection through the downlink with probability density distributions ${p_{AS}}\left( \eta  \right)$ and ${p_{SB}}\left( {\eta } \right)$, respectively, the CM of the two-mode  state at the ground stations for each realization of the transmission factors ${\eta}$  (uplink) and $\eta '$ (downlink) can be constructed by
\begin{equation}
{M_{\eta \,\eta '}} = \left( {\begin{array}{*{20}{c}}
{v\,I}&{\sqrt {\eta \,\eta '} \sqrt {{v^2} - 1} \,Z}\\
{\sqrt {\eta \,\eta '} \sqrt {{v^2} - 1} \,Z}&{\left( {1 + \eta \,\eta '\left( {v - 1} \right) + \chi } \right)\,I}
\end{array}} \right)
\end{equation}
Here, we also assume the QKD protocol is performed in the presence of excess noise variance $\chi$. In realistic implementation of CV QKD over such a scheme, the excess noise can generally come from several sources such as preparation of quantum states at the transmitter, reflection at the satellite, detection at the receiver, excess channel noise, or noise generated by Eve. Here we assume that the excess noise manifests itself only at the receiver and is independent of the fading.

Since ${\eta}$ and $\eta '$ are random variables, the elements of the final CM of the resulting mixed state are calculated by averaging the elements of ${{M _{\eta \,\eta '}} }$ over all possible transmission factors of the two fading channels giving
\begin{equation}
\begin{array}{l}
{M} = \left( {\begin{array}{*{20}{c}}
{v\,I}&{{c}\,Z}\\
{{c}\,Z}&{{b}\,I}
\end{array}} \right)\,, \ {\rm where}
\\ \\
{b} = \int_0^{{\eta _0}} {\int_0^{{{\eta '_0}}} {{p_{AS}}(\eta )\,{p_{SB}}(\eta ')} } \,\left( {1 + \eta \,\eta '\left( {v - 1} \right) + \chi } \right)\,d\eta \,d\eta '\\
\\
{c} = \int_0^{{\eta _0}} {\int_0^{{{\eta '_0}}} {{p_{AS}}(\eta )\,{p_{SB}}(\eta ')} } \,\sqrt {\eta \,\eta '} \sqrt {{v^2} - 1} \,d\eta \,d\eta ' .
\end{array}
\end{equation}
Note that the final state ensemble is a non-Gaussian mixture of the Gaussian states obtained for each realization of ${\eta}$ and $\eta '$.

 Our entanglement-based CV QKD protocols can be performed such that ground station A applies a homodyne  measurement of a mode's quadratures (according to a random bit), or else applies a heterodyne measurement of both quadratures. The ground station B also makes a homodyne measurement of the amplitude or phase quadrature over its mode depending on its own random bit.



 Since the resulting ensemble-averaged state shared by the ground stations is a non-Gaussian state, it cannot be described completely by its first and second moments. Therefore, the key rate we compute based on the CM of the resulting mixed state is essentially based only on the Gaussian entanglement between the terrestrial stations, and therefore  the actual generated key rate may be higher in practice.

In the QKD protocol, Alice and Bob are required to know the channel characteristics, i.e. the channel transmission and the amount of excess noise, in order to bound Eve's information. Since the rate of atmospheric fluctuations are of order kHz, which is at least a thousand times slower than typical transmission/detection rates \cite{fso,20,21}, such channel measurements can be obtained. Note, that in our scheme it is only the \emph{combined} channel transmissivity ${\eta}\eta '$ that is measured at the ground station B.


\section{Comparison of the CV QKD protocols}

We now simulate the performance of our scheme in terms of the estimated key rate.
For all simulations shown, the following assumptions are adopted:
(i) For each simulation, all initial entangled states have the same level of squeezing $r$.
(ii) Beam wander, as modeled by the log-negative Weibull distribution, is used to characterize the two fading channels, with $\beta=1$.
(iii) The two separate fading channels are assumed to be independent, but not necessarily identical.
(iv) The beam wander standard deviations $\,{\sigma _{b\_AS}}\,,\,\,{\sigma _{b\_SB}}$ for the two possible link traversals satisfy $
{\sigma _{b\_SB}} = {k_1}\,{k_2}\,{\sigma _{b\_AS}}$, where $0 \le {k_1} \le 1$ and ${k_2} \ge 0$, respectively, parameterize the beam wander dependence on communication direction and geometries. For clarity the apertures (and beam-spot radii) will be assumed the same at satellite and ground station.
(v) For each CV QKD protocol, Bob  carries out a homodyne measurement on his own component.
(vi) All key rates are calculated in bits per pulse.

\begin{figure}[!t]
    \begin{center}
   {\includegraphics[width=2.9 in, height=2.65 in]{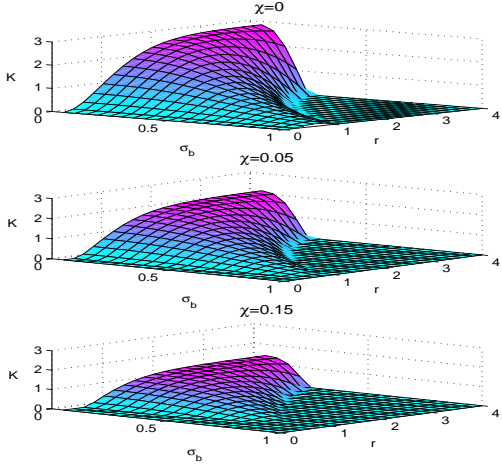}}
    \caption{Estimated key rate $K$ in  a CV QKD protocol between the ground stations where Alice applies a homodyne detection and the RR scheme is implemented with respect to the beam wander standard deviation ${\sigma _b}$ (normalized to $\beta$) in the uplink, and the squeezing level $r$. Here, ${k_1} = 0.4\,,\,\,{k_2} = 0.64$ and $\beta/W =1$.}\label{fig:1}
    \end{center}
\end{figure}

\begin{figure}[!t]
    \begin{center}
   {\includegraphics[width=2.9 in, height=2.7 in]{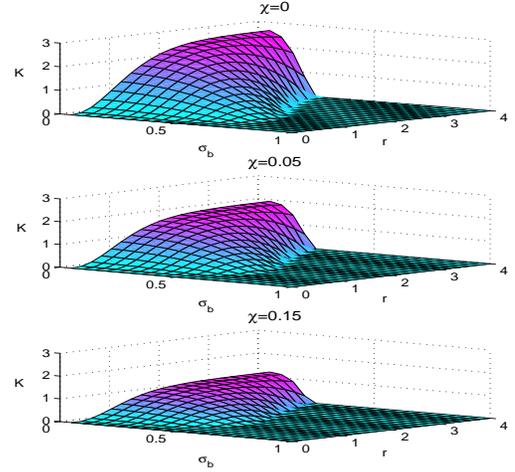}}
    \caption{Same as Fig.~\ref{fig:1} except here a DR scheme is implemented.}\label{fig:2}
    \end{center}
\end{figure}

\begin{figure}[!t]
    \begin{center}
   {\includegraphics[width=2.9 in, height=2.7 in]{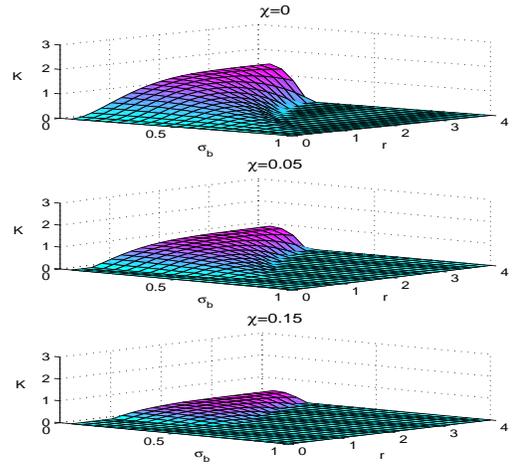}}
    \caption{Same as Fig.~\ref{fig:1} except here Alice applies a heterodyne detection.}\label{fig:3}
    \end{center}
\end{figure}

Fig.~\ref{fig:1}  shows the estimated key rate resulting from the direct QKD scheme in which Alice applies a homodyne detection in the RR scenario, Fig.~\ref{fig:2} displays the case of DR with homodyne detection at Alice, while Fig.~\ref{fig:3} corresponds to the protocol where Alice makes a heterodyne detection in the RR scenario.  The key rate is estimated as a function of beam wander standard deviation ${\sigma_b}$ in the uplink from station A, and the squeezing level $r$ of the initial entangled states in the absence and in the presence of the excess noise $\chi$. The parameters shown in Figs.~\ref{fig:1}-\ref{fig:3} correspond to channels with mean losses of approximately 3dB (at $\sigma_b=0.7$) in the uplink. They are used here only to show the trends expected in the FSO channel. Although not directly related to our specific ground-satellite scenario, such losses are typical of  FSO ground atmospheric links of about 1km length \cite {21}, as well as high-altitude-platform to satellite links of the type discussed in \cite{hap}.

It is evident that  an increase in $\sigma_b$ reduces the key rate since the amount of Gaussian entanglement between the ground stations is diminished by increasing beam fluctuations variance, while increasing the input squeezing is able to partly compensate the fading since the initial entanglement increases. However, for a large squeezing levels at large $\sigma_b$ we see the resulting key rate degrades  since strongly squeezed states are more sensitive to fading. Note that excess noise at the receiver drastically reduces the key rate such that in the presence of high noise the key rate becomes zero for large values of ${\sigma _b}$ (i.e. the high-loss regime). The other point of these results is that when Alice makes a heterodyne detection of (Fig.~\ref{fig:3}) on her part, the key rate is reduced by roughly $50\% $ compared to homodyne detection (Fig.~\ref{fig:1}). For the DR case of Fig.~\ref{fig:2} we find similar results to the RR case of Fig.~\ref{fig:1}, except that the key rate always disappears for losses above a specific threshold. Explicitly we find  in the DR case, the key rate is always zero for values of ${\sigma _b} > 0.7$, which is in agreement with the fact that for fixed attenuation channels, DR protocol only works for losses smaller than 3dB \cite{RR2}.




Although reverse reconciliation is able to improve the key rate at high losses, it is still not sufficient for ground-to-satellite communications which undergo much stronger losses than those illustrated in Figs.~\ref{fig:1}-\ref{fig:3}. Single FSO uplink  ground-to-satellite channels are anticipated to have losses of order 25dB and beyond \cite{fso}. Under such losses, generation of a quantum key will be a fruitless endeavor without use of a highly-selective post-selection strategy.

\section{Post-selection}
In order to enhance the quantum key rate between the ground stations, we apply a post-selection strategy where a subset of the channel transmittance distribution, with high transmittivity, is selected to contribute to the resulting post-selected state used for the quantum key generation. The post-selection strategy which occurs at the receiving ground station is based on classical measurements of the channel transmittance. This strategy has been previously exploited in \cite{21} for a CV QKD protocol over a small-scale single point-to-point fading channel.

\begin{figure}[!t]
    \begin{center}
   {\includegraphics[width=3.6 in, height=1.4 in]{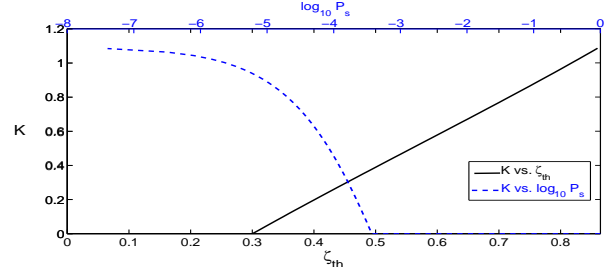}}
    \caption{Estimated key rate $K$ (in bits per pulse selected) in a CV QKD protocol between the ground stations where Alice applies a homodyne detection and the RR scheme is implemented in terms of PS threshold $\zeta _{th}$ (solid line), and success probability of PS ${P_s}$ (dashed line). Here,  $r=1.5$,  $\beta /W = 1,\,\,{\sigma _{b\_AS}}\, = 22\beta ,\,\,{\sigma _{b\_SB}}\, = 2\beta,\,\,\chi=0.15 $. This channel corresponds to a mean loss of 30dB in the uplink, and 10dB in the downlink.}\label{fig:HL1}
    \end{center}
\end{figure}
\begin{figure}[!t]
    \begin{center}
   {\includegraphics[width=3.6 in, height=1.4 in]{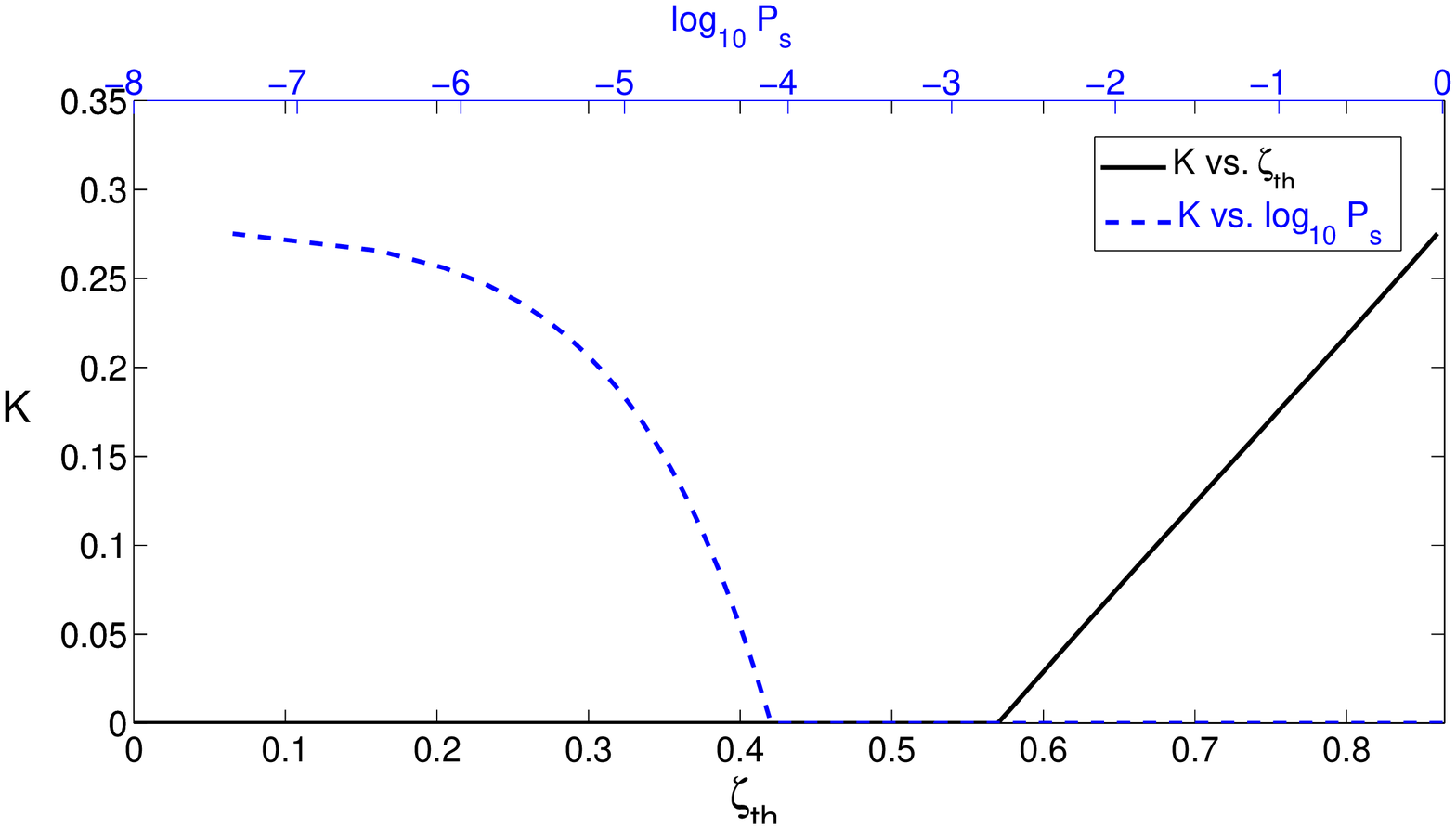}}
    \caption{Same as Fig.~\ref{fig:HL1} except here Alice applies a heterodyne detection.}\label{fig:HL2}
    \end{center}
\end{figure}

For this form of post-selection to operate in our scheme, in addition to quantum information, a large number of coherent (classical) light pulses are sent through fading uplink and then reflected off the satellite in order to measure the transmittance of the combined channel $\zeta  = \eta \,\eta '$ at the receiving ground station, where again $\eta $ and $\eta '$ are random variables describing transmission factors of the uplink and downlink, respectively. The received quantum state is kept or discarded, conditioned on the classical measurement outcome being larger or smaller than a post-selection threshold ${\zeta _{th}}$. Providing we have a form for the probability density distribution $p(\zeta)$, the resulting post-selected CM can be calculated as
\begin{eqnarray}\label{PS2}
\begin{array}{l}
{M^{ps}} = \left( {\begin{array}{*{20}{c}}
{\,v\,I}&{\,{c^{ps}}Z}\\
{{c^{ps}}Z}&{\,{b^{ps}}I}
\end{array}} \right) , \ {\rm where} \\
\\
{b^{ps}} = \frac{1}{{{P_s}}}\int_{{\zeta _{th}}}^{{\eta _0}{{\eta '_0}}} {p(\zeta )} \,\left( {1 + \zeta \left( {v - 1} \right)+ \chi} \right)\,d\zeta \,\\
\\
{c^{ps}} = \frac{1}{{{P_s}}}\int_{{\zeta _{th}}}^{{\eta _0}{{\eta '_0}}} {p(\zeta )} \,\sqrt \zeta  \sqrt {{v^2} - 1} \,\,d\zeta \, .
\end{array}
\end{eqnarray}
Here, ${P_s}$ is the total probability for the combined channel transmission to fall within the post-selected region, and is given by ${P_s} = \int_{{\zeta _{th}}}^{{\eta _0}{{\eta '_0}}} {p(\zeta )} \,d\zeta \, .$
Using ${M^{ps}}$, the key rate emerging from the post-selected entangled state can be computed.  In the high-loss ground-to-satellite scenario we are considering one could expect typically 25-30dB loss in the uplink and 5-10dB in the downlink. Fig.~\ref{fig:HL1} and Fig.~\ref{fig:HL2} show  expected key rates in such losses. In Fig.~\ref{fig:HL1} the key rate is calculated for the case where Alice makes a homodyne measurement in the RR scenario in the presence of noise. Fig.~\ref{fig:HL2} is identical except that Alice makes a heterodyne measurement. The key in both figures is illustrated with respect to the post-selection threshold ${\zeta _{th}}$ and success probability ${P_s}$. The figures explicitly show the trade-off in increased key rate (as the threshold value increases) at the cost of lower success probability. Note that in these calculations no closed-form solution for $p(\zeta)$ could be used, so a numerically determined form was utilized. It is important to realize that these are key rates per \emph{post-selected pulse}. That is they are the key rates determined only from the final ensemble of post-selected states. As such, the input pulse rate at the sender  must be multiplied by the post-selection probability $P_s$, and the key rate $K$ (bits per pulse selected), in order to find the final key rate in bits-per-second. Note, significantly higher rates than those illustrated in Figs.~\ref{fig:HL1}-\ref{fig:HL2} can be achieved if entangled states are generated in the satellite directly, as we discuss next.




\section{Discussion}
Another approach to entanglement-based CV QKD distribution is through on-board generation of entangled pairs within the satellite itself. In this alternative scheme one of the entangled modes is sent directly to station A with the other mode sent directly to station B. Although such a scheme increases the complexity at the satellite it does have the advantage of having no uplink channels.
For LEO satellites one could expect losses in downward links to be better than the losses in upward links by levels of order 20dB, e.g. \cite{r7}. From an application of the RR performance analysis given in \cite{weed2} to this lower-loss fading scenario, we find  the alternative on-board generation scheme generates $K=0.83$ at $P_s=10^{-3}$ (at $\zeta _{th}=0.8$). Relative to the direct QKD scheme  of Fig.~\ref{fig:HL1}, a key rate $K=0.83$ at $P_s=10^{-3}$ would represent an approximately $100$ fold increase in the bits-per-second final key rate, thereby illustrating the trade-off in performance versus (satellite-based) complexity.

Possibilities for improving the direct QKD scheme are provided by multiple-beam technology (spatial diversity) as applied to the FSO scenario \cite{mimo}.
In the direct QKD scheme, an optimal diversity gain in the generated quantum key rate will require some form of quantum coding across the beams in the uplink  - a sophisticated quantum-engineered task. However, simpler no-coding diversity set-ups will still significantly increase the success probability of post-selection and a corresponding increase in the key rates.
The remaining engineering complexity in these latter set-ups lies largely in the integration of  beam selection at the sender and receiver (which may be meters apart on the ground), and in the reflection of multiple beams  at the satellite.

Note again, that the CV QKD rates presented here are based on the assumption of an infinite number of signals being sent between the sender and receiver.  Of course, in reality all QKD deployments undergo only finite signalling. However, such finite signalling effects are
 of particular relevance to our direct QKD scheme  due to the highly selective nature of our post-selection strategy.  As such, the rates determined here can only be described as indicative of future performance if the effects of finite signalling can be shown to be negligible. Security proofs based on finite signals are difficult but progress has been made recently, e.g. \cite{Furr0} \cite{sec1} \cite{Furr}. If future theoretical studies\footnote{ We  note that classical satellite communications are subject to both special and general relativistic corrections, e.g. \cite{ashby}. Relativistic effects on satellite-based quantum communications have also been investigated recently, with small but observable effects predicted \cite{david}.  In this work the small impact  such relativistic corrections may have on key rates are ignored. However, it would be useful if future satellite-based CV QKD security proofs accounting for finite signalling effects also formally couple-in all relevant relativistic effects.}  could find that under the large losses associated with ground-to-satellite fading channels, a total signaling number of order $10^{10}$ negates any significant finite-size effects then all the results provided in Figs.~\ref{fig:HL1}-\ref{fig:HL2} would be immediately applicable. Improvements in the input pulse rate (typically $10^8$Hz), the addition of long-term quantum memory, use of multiple satellites (or multiple pass-overs), and/or use of multiple beams, would drive downwards the difficulty in realizing a QKD system in which finite signalling effects could be ignored.
\section{Conclusions}
In this work we have explored a quantum communication architecture based on reflection from a LEO satellite in order to perform Gaussian entanglement-based CV QKD.
Utilizing reverse reconciliation in the post-processing strategy combined with a highly-selective post-selection strategy we have found that a useful quantum key rate   may be  achievable.
The results given here represent the first quantitative assessment of CV QKD via reflection off a LEO satellite, and provide confidence that an experimental validation of space-borne CV QKD is within reach.
%

\end{document}